\documentclass[aps,pre,superscriptaddress,twocolumn,showpacs]{revtex4-1}

\usepackage{amsfonts}
\usepackage{amsmath}
\usepackage{amssymb}
\usepackage{amstext}
\usepackage{amsthm}
\usepackage{graphicx}
\usepackage{dcolumn}
\usepackage{bm}
\usepackage{color}
\usepackage{float}

\begin{document}

\title{Uncovering functional brain signature via random matrix theory}

\author{Assaf Almog}
\affiliation{The Big Data Lab, Department of Industrial Engineering, Tel-Aviv University (Israel)}
\affiliation{Lorentz Institute for Theoretical Physics, University of Leiden (The Netherlands)}

\author{M. Renate Buijink}
\affiliation{Laboratory for Neurophysiology, Department of Molecular Cell Biology, Leiden University Medical Center, Leiden (The Netherlands)}

\author{Ori Roethler}
\affiliation{Laboratory for Neurophysiology, Department of Molecular Cell Biology, Leiden University Medical Center, Leiden (The Netherlands)}

\author{Stephan Michel}
\affiliation{Laboratory for Neurophysiology, Department of Molecular Cell Biology, Leiden University Medical Center, Leiden (The Netherlands)}

\author{Johanna H. Meijer}
\affiliation{Laboratory for Neurophysiology, Department of Molecular Cell Biology, Leiden University Medical Center, Leiden (The Netherlands)}

\author{Jos H. T. Rohling}
\affiliation{Laboratory for Neurophysiology, Department of Molecular Cell Biology, Leiden University Medical Center, Leiden (The Netherlands)}

\author{Diego Garlaschelli}
\affiliation{NETWORKS Unit, IMT School of Advanced Studies, Lucca (Italy)}
\affiliation{Lorentz Institute for Theoretical Physics, University of Leiden (The Netherlands)}

\date{\today}

\begin{abstract}
The brain is organized in a modular way, serving multiple functionalities. This multiplicity requires that both positive (e.g. excitatory, phase-coherent) and negative (e.g. inhibitory, phase-opposing) interactions take place across brain modules. Unfortunately, most methods to detect modules from time series either neglect or convert to positive any measured negative correlation. This may leave a significant part of the sign-dependent functional structure undetected. Here we present a novel method, based on random matrix theory, for the identification of sign-dependent modules in the brain. Our method filters out the joint effects of local (unit-specific) noise and global (system-wide) dependencies that empirically obfuscate such structure. The method is guaranteed to identify an optimally contrasted functional `signature', i.e. a partition into modules that are positively correlated internally and negatively correlated across. The method is purely data-driven, does not use any arbitrary threshold or network projection, and outputs only statistically significant structure. In measurements of neuronal gene expression in the biological clock of mice, the method systematically uncovers two otherwise undetectable, negatively correlated modules whose relative size and mutual interaction strength are found to depend on photoperiod. The neurons alternating
between the two modules define a candidate region of functional plasticity for circadian modulation.
\end{abstract}

\maketitle

\newpage
\section{Introduction}
Understanding how billions of neurons collectively self-organise into a functionally ordered brain able to coordinate a variety of neural, cognitive and bodily processes is probably the most fundamental quest in neuroscience.
Over the last decades, evidence has accumulated that the functional organisation of the brain is modular and hierarchical \cite{Modular}. This means that the brain appears to be partitioned into mesoscopic `functional modules' where each module is composed of neurons with a relatively similar dynamical activity, while different modules are comparatively less related to each other. Each such module may also contain submodules hierarchically nested within it.

Reliably identifying functional modules is a nontrivial task because of their irreducibility to contiguous anatomical regions defined \emph{a priori} and/or to local neighbourhoods in the underlying structural network of neuron-to-neuron anatomical connections \cite{neuro}.
Indeed, while on the one hand functional modules partly reflect the local brain anatomy, on the other hand major deviations between functional and structural networks are observed. 
One key example is the distinctive `long-range' left-right splitting of some functional modules: often, a single module is found to be composed of two or more spatially non-contiguous populations of neurons, located in possibly distant (sometimes symmetric, sometimes asymmetric \cite{Corballis}) regions in the left-right direction \cite{Nicosia,Michel}. 
As an opposite example, an anatomically well-defined brain region can be functionally heterogeneous \cite{Antle,Buijink} and sometimes even display anti-correlation between the activity of some of its parts \cite{Iglesia,Ohta}. 
These examples indicate the lack of a one-to-one correspondence between structural and functional modules, showing that it is in general impossible to infer the latter purely from spatial information. 
Indeed, it is expected that the mapping between functional and structural networks is many-to-one, thus allowing a diversity of functions to arise from a common neuronal anatomy \cite{neuro}. 
On top of this, both structural and functional brain networks are characterized by \emph{plasticity}, i.e. possibility of temporal rearrangements, but at typically different spatial and temporal scales.

\begin{figure}
\centering
\includegraphics[width=1\linewidth]{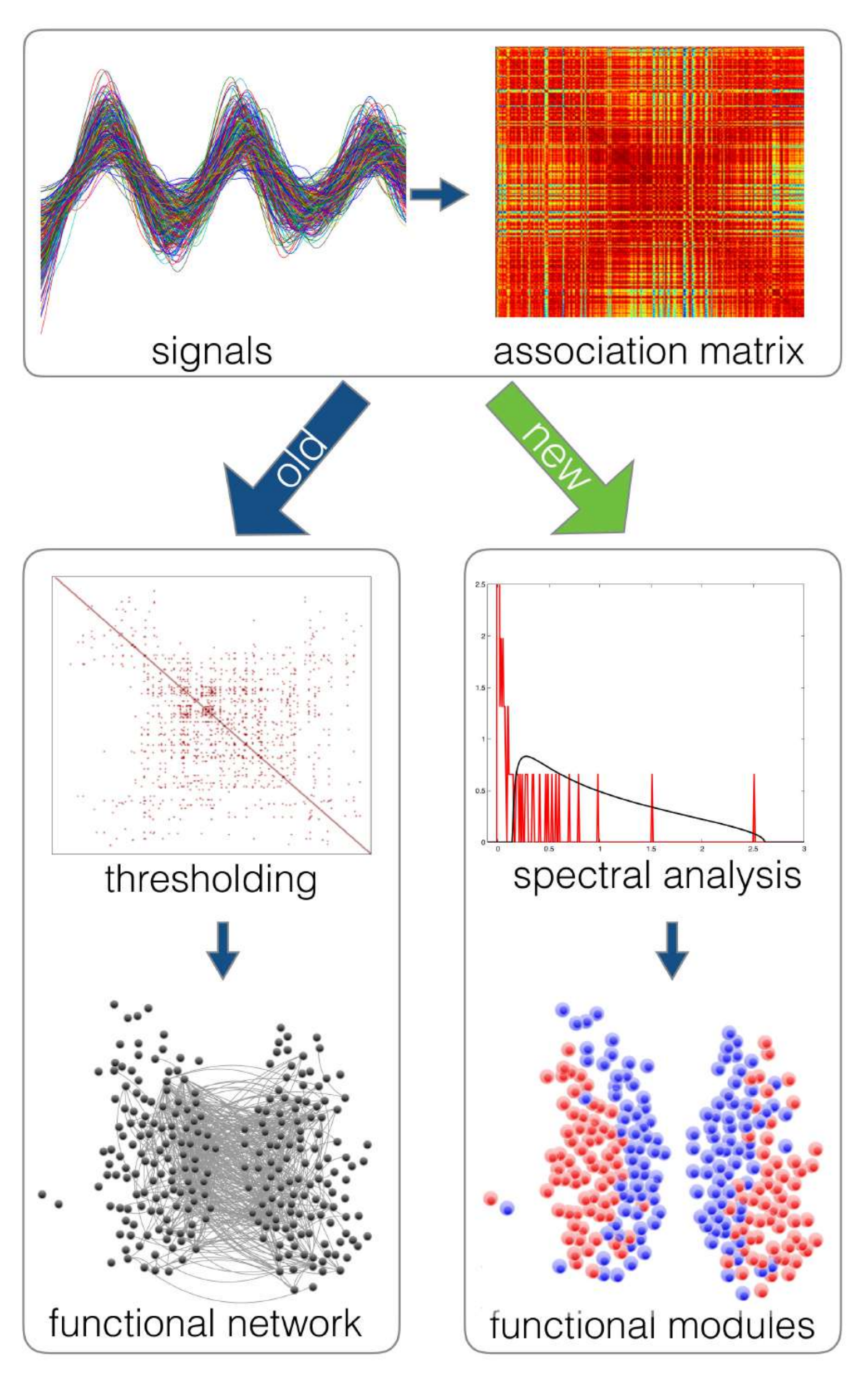} 
\caption{Illustration of the procedure of functional module identification from time series data (top) in the stardard approach (bottom left) and in our method (bottom right). 
In the standard approach, an arbitrary threshold is defined and the original matrix is projected onto a functional network. This comes at the price of discarding the majority of the data, most notably the negative correlations, and makes the output threshold-dependent. Moreover, modules are searched for in the projected network using null models that are valid for graphs with independent edges, but not for correlation matrices. 
In our method, we compare the empirical correlation spectrum against a null model specifically tailored for correlation matrices. This produces a filtered correlation matrix that is subsequently searched for modules. These modules are guaranteed to be statistically significant, noise-free, overall positively correlated internally and overall negatively correlated across.
By directly producing a partition of the original time series into modules, our method bypasses the functional network projection, avoiding the use of a threshold.}
\label{fig:intro}
\end{figure}

Precisely because they cannot be reduced to `spatially obvious' brain regions, functional modules must entail an emergent, non-structural level of neural organisation which can only be investigated via the explicit analysis of time series of activity of individual neurons or, at a more coarse-grained level, regions of interest (ROIs). 
More specifically, recordings of multiple time series are normally used to construct an association (e.g. cross-correlation, mutual information, etc.) matrix capturing the mutual relations between pairs of ROIs (see Fig~\ref{fig:intro}).
Next, the matrix can be analysed in different ways to detect the presence of functional dependencies or structure in the system.

Importantly, these dependencies can be positive ($+$) or negative ($-$), leading to measured correlation or anti-correlation.
For instance, synaptic interactions between neurons will influence their mutual phases and lead to different states of synchronization in a brain circuit. The degree of synchronization ($+$) versus desynchronization ($-$) is important for neural function and a disturbance in this balance can contribute to neurological disorders. 
In the paradigmatic example of the central mammalian clock situated in the \emph{suprachiasmatic nucleus} (SCN) of the hypothalamus, the state of synchronization of neurons can influence responses of the circadian system to light and is actually used to encode seasonal changes in day length. It has been suggested that inhibitory ($-$) as well as excitatory ($+$) neuronal interactions will contribute to the phase differences observed under different photoperiods \cite{Myung,DeWoskin}. The balance between excitatory and inhibitory activity (E/I balance), which is a hallmark of healthy network performance, can actually change with photoperiod \cite{Farajnia}. 

The motivation for the present paper is the expectation that, in the brain and in possibly many other biological networks as well, the presence of both positive and negative interactions should have a significant impact on how the modular functional organization is both mathematically defined and empirically identified. 
For instance, even within a functionally homogeneous region there may be negatively correlated substructures arising from the need to create and/or modulate the internal mutual phase relationships. Similarly, across two functionally distinct modules there may be a need for dependencies of both negative and positive sign, depending on whether the two functions should inhibit or enhance each other.
Consequently, we stress that a proper definition of functional modules should take the sign of the defining correlations into serious account and tools should be devised to reliably identify such sign-dependent structure from time series data. This is crucial in order to map how function is distributed across the modular brain landscape and to properly constrain models of the underlying neural dynamics.

In this paper, we argue that the available approaches to the theoretical definition and empirical detection of functional modules treat negative dependencies in essentially unsatisfactory ways.
On one hand, most techniques either entirely dismiss negative values or turn them into positive ones, thereby using no information about the sign of the dependency.
On the other hand, the few methods that do take negative correlations into account use (null) models that treat all pairwise correlation coefficients as statistically independent entities, thus violating important structural properties of correlation matrices. 
Other popular approaches like Principal Component Analysis (PCA) or Independent Component Analysis (ICA) look for \emph{independent}, rather than anticorrelated, components, thus serving a different purpose.
Moreover, most of these approaches fail to provide a stastistical validation of the modules identified, and are therefore prone to misidentification due to the presence of both ROI-specific noise and brain-wide common trends obfuscating the underlying mesoscopic modular patterns.

Here, we propose a novel method that targets specifically the positive and negative interactions in brain data and filters the underlying noise and common trends using an appropriate null model based on Random Matrix Theory (RMT).
Our approach generalizes a recent community detection method tailored for correlation matrices \cite{MacMahon,Almog}, originally formulated for financial time series that have an inherently random and non-periodic pattern, and extends it to the case where arbitrarily structured temporal trends are allowed.  We also pay specific attention to the fact that noise and global trends have a previously overlooked coupled effect on the spectrum of correlations, and we rigorously correct for this coupling.
Technically, the method makes use of a modified Wishart ensemble of random correlation matrices constructed using precisely the same common trend and expected noise level as the empirical time series, but under the null hypothesis that no modular organization is present.
This ensemble serves as a natural, reliable and more appropriate null model for correlation matrices arising in brain research. 
A comparison between empirical and null correlation matrices reveals the functional modules present in the data and by construction absent in the model. 

The resulting method is threshold-free and does not require the arbitrary projection onto a network (see Fig~\ref{fig:intro}).
Moreover, in contrast with most of the current approaches, it is designed to yield an optimally sign-contrasted structure, where positive interactions are clustered inside the modules and negative values are expelled across modules. We call the resulting optimized structure the \emph{functional signature} of the system.
This structure is composed of functional modules whose overall internal correlation is guaranteed to be positive and whose overall mutual correlation is guaranteed to be negative. 
The method only outputs statistically significant structure, if present.
We should stress that in any stage of the process there are no presumptions about the output of the method (such as a predefined number or size of modules) and the results are completely and non-parametrically driven by the data themselves.  
If needed, the method can be used iteratively to detect sub-modules hierarchically nested within modules.

Besides formulating the method, we apply it to the analysis of the aforementioned SCN, which is responsible for regulating the circadian rhythms of physiology and behaviour in mammals. We chose the SCN of mice because of its relatively small size (ca 20,000 neurons) and high degree of functional plasticity. 
Single SCN neurons are capable of generating circadian rhythms in, amongst others, gene expression and electrical activity. The phase differences between the cells can vary with changes in the environment, such as different photoperiods or prolonged light exposure, or with an attenuation of the degree of coupling between the neurons as seen in aging or disease.
This makes the SCN an optimal case study for a dynamic network of neurons with different internal oscillations, mechanistically coupled to E/I processes. 

We show how our method can be used to reliably search the SCN for sign-dependent functional modules reflecting the phase ordering of oscillating cell populations, based on both strength and sign of their coupling interactions.  
We use samples taken from mice that were subjected to different photoperiods.
The method identifies two otherwise undetectable clusters of functionally connected SCN neurons that have a strong resemblance to a known core/shell distinction \cite{Morin} and that have never been found before without the use of prior knowledge.
Importantly, we are able to detect physiological differences present in different photoperiods in the functional signature of the two clusters.
We find that the sizes of the two modules change with photoperiod as the result of a majority of neurons remaining in the same module irrespective of photoperiod, and a minority alternating between the two modules at their interface.
This finding highlights a possible population of alternating neurons responbile for the functional plasticity required for adjustment to photoperiod and circadian modulation.

\begin{figure*}{}
\centering
\includegraphics[width=1\linewidth]{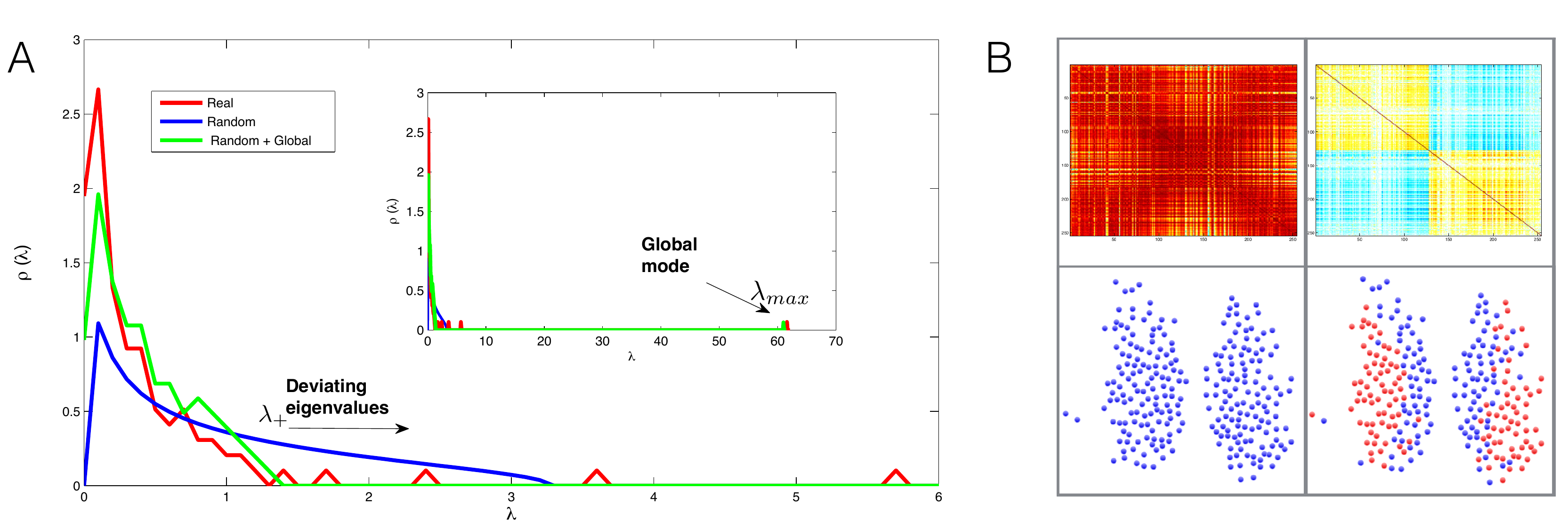} 
\caption{ (A)
Empirical eigenvalue density versus calculated eigenvalue density for the two random models. (B)  
The community structure of the SCN as resolved by our method.
On the left, is the community structure detected by random model without filtering the global mode (Random).
On the right, is the community structure detected by random model once the global mode is filtered (Random + Global).
In the bottom panels are the partitions detected, where each community is marked with a different colour.
In the top panels are the corresponding resolved filtered correlation matrices displaying the resolved structure as a block matrix.}
\label{fig:result}
\end{figure*}

\begin{figure}{}
\centering
\includegraphics[width=0.9\linewidth]{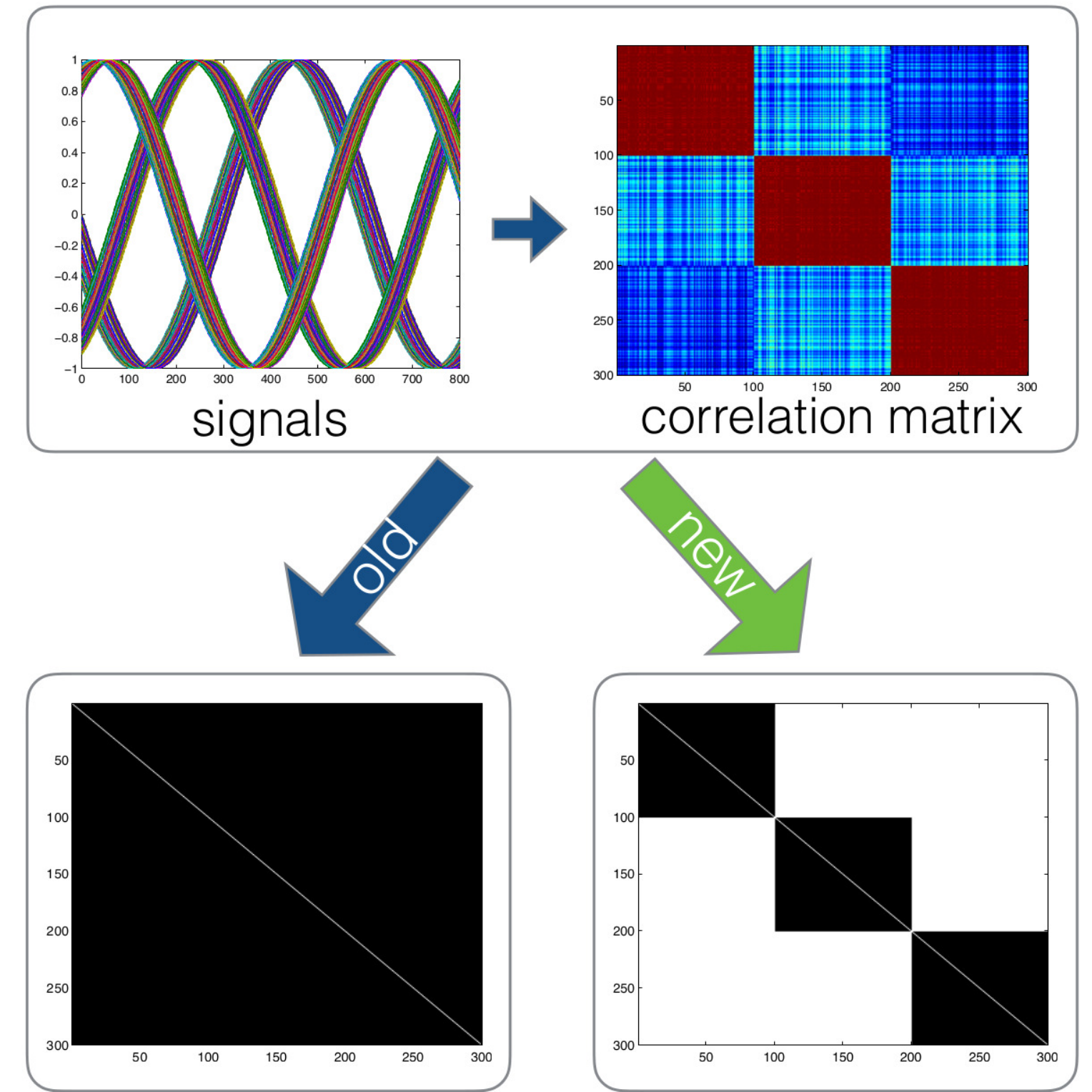} 
\caption{Illustration of the `merging bias' in a comparison between the method by Rubinov and Sporns \cite{SpornsSigned} (based on a null model with independent entries of the correlation matrix) and our alternative approach (based on the more appropriate null model with dependent entries constructed from random matrix theory). Top: 300 synthetically generated time series in a system with 3 modules, each containing 100 oscillating serie with random phases (left) and the corresponding correlation matrix showing a clear block structure (right). Bottom: output of the Rubinov-Sporns method (left) and our method (right) in terms of likelihood matrices indicating the frequency with which two neurons are found in the same community in 1000 runs of both methods. We can see that the Rubinov-Sporns method suffers from merging bias and clusters all the signals into one module (bottom left), while our method correctly separates the 3 modules (bottom right).}
\label{fig:Compare}
\end{figure}

\begin{figure*}{}
\centering
\includegraphics[width=0.99\linewidth]{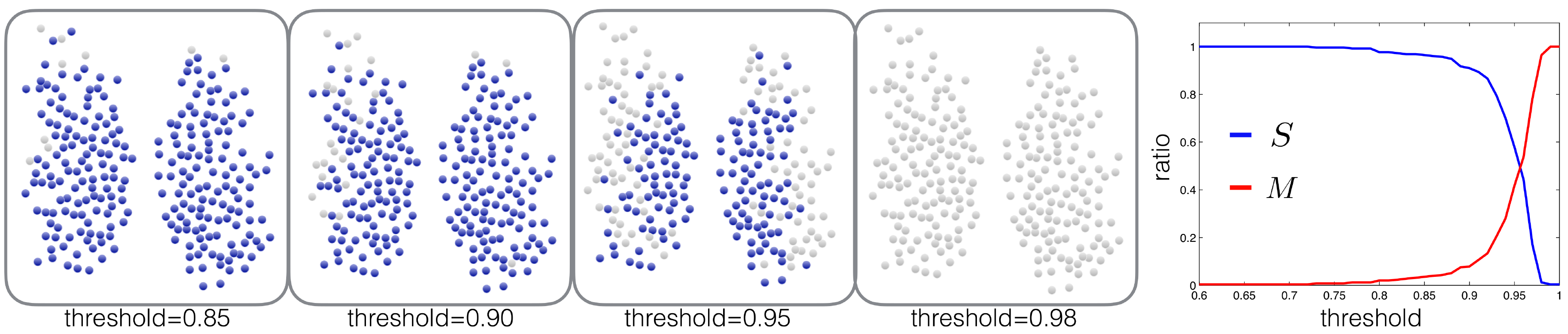} 
\caption{The community structure of the SCN as resolved by a standard threshold approach. On the left, we plot the community structure, resolved by the standard method, for different thresholds. In blue are the nodes that belong to the large cluster, while in gray are isolated nodes ('communities' that only contain one node). 
In the right panel, we plot the fraction of nodes in the largest connected component $S$ in blue, and the fraction of communities detected $M$ in red.}
\label{fig:standard}
\end{figure*}

\section*{Results}

\subsection*{Limitations to overcome in the identification of sign-dependent functional modules}

Our approach aims at overcoming various limitations of the existing methods. It is therefore convenient to briefly mention these limitations in order to gradually introduce some of the defining elements of our method.

First, we want to avoid the use of thresholds on the entries of the correlation matrix.
Indeed, most approaches identify functional modules via the introduction of a threshold used to project the original correlation matrix into a network (see Fig~\ref{fig:intro}) \cite{FUNE,Rubinov}. On this network, various graph-theoretic quantities can be measured to identify modules in terms of e.g. connected components \cite{Palla}, rich clubs \cite{Colizza}, $k$-cores \cite{Seidman} or communities \cite{Fortunato}.
The well known limitations of this approach are the uncontrolled information loss induced by discarding some of the observations, the complete arbitrariness of the choice of the threshold value, and the resulting unavoidable threshold-dependence of the output \cite{Garrison}.
Moreover, since thresholds are introduced to project the original matrix into a sparse network, and since the number of negative entries in such matrix is usually smaller than that of positive ones, this procedure essentially imposes a positive threshold, thereby completely disregarding all the negative correlations.

Second, we want to avoid turning the negative correlations into positive ones.
Based on the (correct) consideration that negative correlations indicate functional dependency (rather than no dependency), many approaches aim at exploiting both positive and negative values as cohesive interactions in the definition of functional modules.
To this end, they take e.g. the absolute value or the square of the original correlations. 
However in this way the negative correlations are treated just like the positive ones, making it impossible for the output modules to encode any information about the original sign of the functional dependencies.
We instead believe that the sign should be retained and used as a repulsive interaction in the definition of modules, with the understanding that the latter should not be interpreted as functionally independent of each other, but rather as dependent sub-modules in mutual anticorrelation, possibly nested within larger modules that may eventually be functionally unrelated.

Third, we want to avoid the `merging bias' that affects even the few remaining methods that do preserve the sign of correlations in the definition of modules \cite{SpornsSigned,Arenas,Traag}. 
These methods are adaptations of the so-called `modularity maximization' techniques introduced in the literature about community detection in networks and targeted at finding groups of nodes that are more densely connected internally, and less densely connected across, than expected under a random null model \cite{Fortunato}. 
The main null models for networks have statistically independent links, i.e. a link can be placed between any two nodes without affecting the probability of placing links elsewhere in the network. 
The methods that generalize these null models to correlation matrices extend them in the direction of allowing links with both positive and negative weight, but unfortunately retain the assumption of independent matrix entries \cite{SpornsSigned,Arenas,Traag}. 
While justified for networks, this assumption becomes incorrect for correlation matrices, whose entries are subject to basic `metric' properties that make them depend on each other \cite{MacMahon}. 
For instance, negative triangular relationships of the type $C_{i,j}<0$, $C_{j,k}<0$, $C_{k,i}<0$ are in general very rare in empirical correlation matrices (and become impossible if $C_{i,j}=C_{j,k}=C_{k,i}=-1$), while they are much more likely in a null model with independent entries.
This effectively creates the systematic bias of erroneously interpreting the absence or scarcity of such negative triangles in the data as strong statistical evidence for the nodes $i$, $j$ and $k$ being `attracting' each other.
As a net result, the three nodes are likely to be merged in the same module (hence the merging bias), although their mutual anticorrelation represents statistical evidence that they should in fact belong to three separate modules.

Fourth, we want to avoid misidentification due to the presence of common trends across all ROIs in the sample. 
Indeed, depending on the spatial and temporal resolution of the data, experimental time series may contain a multitude of periodic or systematic trends at different frequencies (e.g. heartbeat, breathing, circadian rhythms) that impart an overall positive correlation to all or several ROIs, without actually representing any real functional relatedness among the ROIs themselves.
One of the side effects of such `global mode' is a reduction of the detectability of the underlying modular structure. 
Certain techniques aim at solving this problem by preliminarily subtracting the measured average trend from each time series separately (thus effectively removing the global mode), and then calculating the resulting correlation matrix. 
This procedure has been criticised because it tends to generate both positive and negative correlations by construction, with no guarantee that the corresponding signs represent a true signature of functional modularity, e.g. even if the original time series were all independent and their increments relative to the average trend were merely due to chance or noise.

Fifth, and connected to the point above, we want to accurately characterize the level of noise in the data. This point is connected to many of the points above. For instance, being able to separate noise from information would allow us to avoid the use of arbitrary thresholds, discriminate between true and random modularity, and arrive at a safer definition of modules based on trends relative to the global one, thus enhancing the detectability of functional substructure.


\subsection*{A random matrix null model for correlation matrices of neural activity}

We are now ready to introduce our method which is designed in order to avoid the limitations described above.

Given an empirical correlation matrix constructed from multiple time series of neuronal activity, our method looks for functional modules upon removing the joint effects of noise in the data and of common temporal trends, as both features may obfuscate the empirical identification of possible underlying substructure.
For this task the method first introduces a null model that serves as a random benchmark, thus accurately highlighting the non-random modular patterns in the empirical correlation matrix.  
This improved null model, based on random matrix theory, takes into account cell to cell variability and does not require the unrealistic assumption that the time series are stationary.
Therefore we can allow for any temporal modulation [see Fig S1], both in individual time series and in their resulting common trend. This is very important, given the strongly time-dependent nature of functional brain data in general, and of our time-modulated oscillating signals in particular.
So, even if the calculation and interpretation of correlation matrices usually assumes stationarity, here we can statistically treat correlation matrices calculated from nonstationary data as well. 

The first step is an exact calculation of the combined, undesired effects of noise and common trends on the density of eigenvalues $\rho(\lambda)$ of a theoretical cross-correlation matrix. 
This step corresponds to the definition of a null model for a correlation matrix without modular patterns, but with a noise level calibrated to the observed one and with a global trend that exactly follows the one in the empirical time series. The output of this first step is illustrated in Fig \ref{fig:result}A. The density of eigenvalues, which is calculated exactly in the null model [see SI], features one largest eigenvalue $\lambda_{\rm max}$ due to the global trend, plus a ``random bulk'' extending between a minimum ($\lambda_-$) and a maximum ($\lambda_+$) eigenvalue. 

The second step is a filtering of the original correlation matrix via the identification of the empirical eigenvalues that deviate, in a statistically significant manner, from the ones predicted by the module-free null model. 
In practice, this reduces to the selection of the empirical eigenvalues that are found in the range $(\lambda_+,\lambda_{\rm max})$. 
A crucial result in this study, overlooked in previous analyses \cite{MacMahon}, is a precise calculation of $\lambda_+$ showing that the higher $\lambda_{\rm max}$, the lower $\lambda_+$. The fact that the values of $\lambda_{\rm max}$ and $\lambda_{+}$ depend on each other is a proof that noise and global trends jointly affect the features of the expected eigenvalue density of the correlation matrix. Our calculation of $\lambda_+$ allows us to recover statistically significant features of the empirical correlation matrix that would otherwise be incorrectly classified as noise.
Looking again at Fig.~\ref{fig:result}, we indeed see the presence of eigenvalues in the empirical spectrum (red) that deviate from our adjusted null model (green) and include eigenvalues that would be incorrectly classified as noisy if $\lambda_+$ were not corrected for $\lambda_{\rm max}$ (blue). This step is completed by the selection of the eigencomponent of the correlation matrix associated with the deviating eigenvalues.
The resulting, cleaned component of the original matrix contains statistically significant, noise- and trend-filtered information about the presence of functional modules. 

\subsection*{Detecting functional signature in neural systems}

Once the original correlation matrix has been filtered by the null model, only the statistically significant dependencies are guaranteed to remain in the matrix.
At this point our aim is the identification of functional modules that are positively correlated internally and negatively correlated externally. This can be transformed into an optimization problem. 
We employ community-detection techniques that take the filtered correlation matrix as input and return the optimized partition of the system into functional modules. 
The optimized partition will tend to place the positive dependencies (correlation) inside the clusters while expelling the negative dependencies (anti-correlation) across the clusters. We should stress that, by construction, the emergent functional structure will be detectable only if it is statistically significant.
Moreover, the number of detected clusters is not defined \emph{a priori}, and is found automatically by the method itself.

It should be noted that, while the use of information contained in the eigenvectors of the largest eigenvalues is common to other methods (such as Principal Component Analysis and it generalization, aka Independent Component Analysis \cite{ICE1,ICE2}) as well, our approach distinguishes itself from these approaches in various respects. First, those methods look for the \emph{independent} components in which the orginal signal can be optimally decomposed, while our aim is to pinpoint the \emph{anticorrelated} groups of units. Second, our iterative optimization procedure reformulated for correlation matrices guarantees that the final output is maximally contrasted in terms of the signs of the detected modules. Finally, the other approaches focus on the strongest eigenvalues but do not implement a null model, tailored to capture both local noise and global trends, to assess which of the eigenvalues are informative and which are noisy. Indeed, in ICA the desired number of components has to be specified by the user, whereas in our method the optimal number of modules is given as output by the algorithm.

By using an appropriate null model that does not have independent entries of the correlation matrix, our method avoids the merging bias of other methods described above. To illustrate this, in Fig~\ref{fig:Compare} we show a synthetic sample with 300 oscillating signals divided into 3 main groups, in each of which 100 signals are randomly assigned different phases around a `master signal'.
We can clearly see that due to the differences in phase between the groups, the relations between different groups become negative (anti-correlation). 
We then process the correlation matrix with the (independent-entries) method proposed in \cite{SpornsSigned} and with our method, and the two methods yield significantly different results. 
While our method is able to cluster the 3 groups perfectly, the general modularity method clusters the whole system into one community.
This is the result of the unrealistically homogeneous null model in the latter method, which disregards the statistical dependencies that are present even between correlations of random signals.
We should stress that this limitation is relevant to all the methods that are mentioned in \cite{SpornsSigned}, precisely because of the wrong use of the null model designed for networks and not for time series.

\subsection*{Uncovering the hidden functional signature of the SCN}

The brain region we apply our method to is the suprachiasmatic nucleus (SCN), located in the hypothalamus in the brain, and recognized as the site of the central circadian clock in mammals. This clock is important for the regulation of our daily and seasonal rhythms. It has been shown that the neuronal network organization of the SCN changes in different photoperiods \cite{VanderLeest}, however, the mechanisms behind these changes are still elusive. Furthermore, only a subset of neurons within the SCN network are directly responsive to light \cite{Rohling1}, which poses the question how encoding for seasonally changing day length is achieved in the SCN network. 
The SCN is a prototypical example of a brain structure for which resolving functional organization is challenging for the reasons outlined above: it consists of about 20000 neurons that are spatially close (total size of ${1}$ ${mm}^3$ - so, structurally speaking, these neurons form a single densely connected cluster, whose only anatomical substructure is a left-right split into two lobes) while at the same time displaying a high variability in terms of the signals of the constituent neurons.

Currently, brain networks are most often derived from data acquisition techniques that do not have the possibility to perform recordings at the single cell level. Techniques such as (functional) Magnetic Resonance Imaging ((f)MRI), Electroencephalography (EEG) or Magnetoencephalography (MEG) use brain regions as nodes in the network and fiber bundles between these regions as edges. We investigate the SCN network at the micro-scale where nodes are single cells and edges are functional connections between the cells.
We use single-neuron data on gene expression of a clock gene \emph{period2} in the SCN. The data were sampled every hour for at least three days by means of a bioluminescence reporter PER2::LUC.

We first perform a standard analysis based on the mainstream method [see Fig~\ref{fig:intro}] for detecting communities via functional networks. This is a useful reference as a comparison with our own method. In Figure \ref{fig:standard} we present the community structure, resolved by the standard method, for different thresholds. In blue are the nodes that belong to the large cluster, while in gray are isolated nodes (communities that only contain one node). 
In the right panel, we plot the fraction of nodes in the largest connected component $S=\frac{LCC}{N}$ in blue, and the fraction of communities detected $M= \frac{Communities}{N}$ in red. 
It is evident that applying different thresholds essentially detaches isolated nodes from the large cluster, and there is no optimal value for the threshold.
Therefore, the standard method can only observe a ``radial gradient'' of connectivity, and there is no sense of multiple communities of neurons, which is one of the signatures of functional as opposed to structural connectivity. This poor performance of the method is a known limitation  when  applied to very dense networks.

\begin{figure}{}
\centering
\includegraphics[width=1\linewidth]{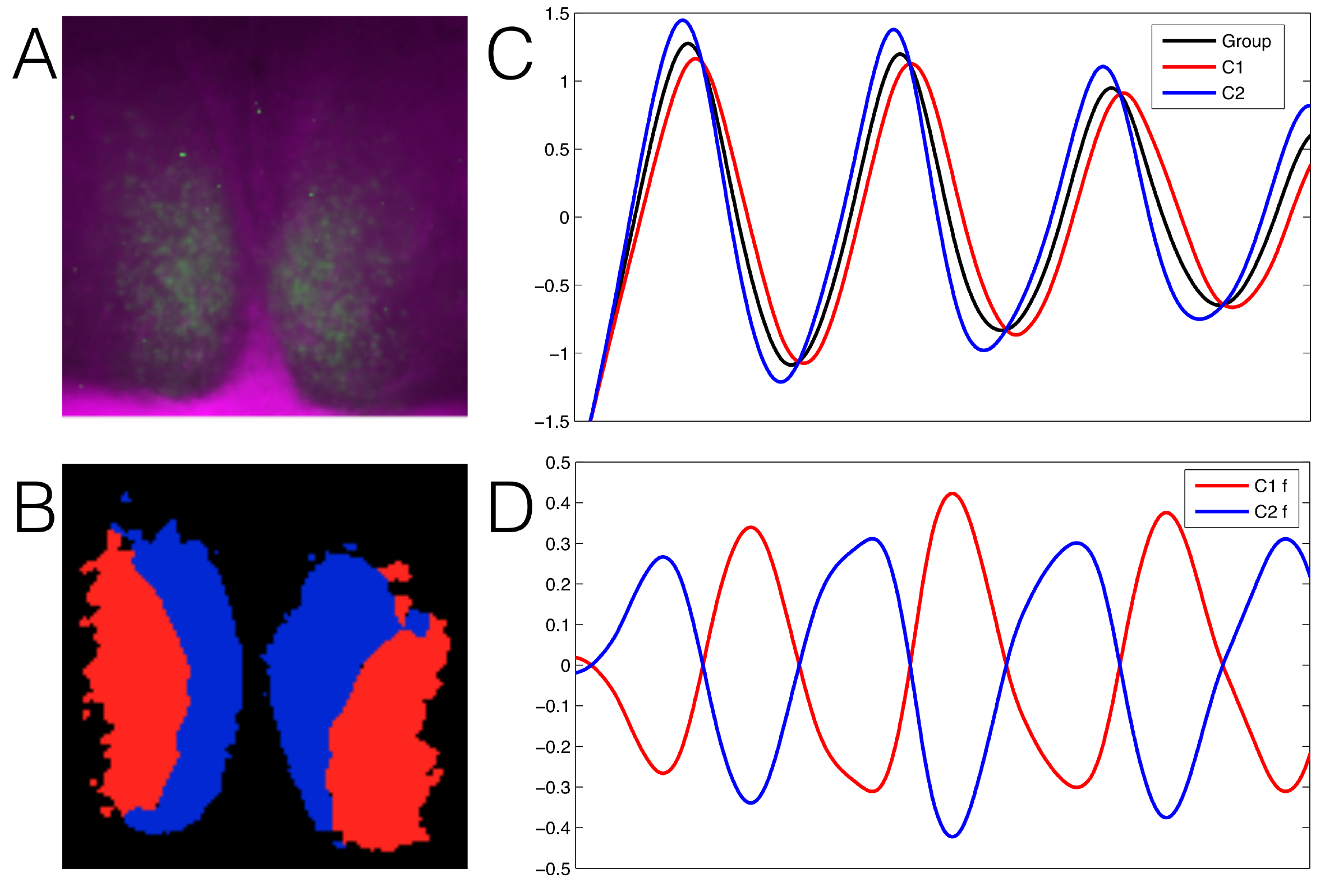} 
\caption{(A) The bioluminescence image of one SCN sample. (B) The plotted average partition over all the samples. (C) The plotted average signal of the whole system (in black) versus the mean signals of the two detected communities (in red and blue) for one SCN sample. (D) The plotted average residual signals of the two communities of one SCN sample, once the global signal is subtracted.
}
\label{}
\end{figure}

\begin{figure*}{}
\centering
\includegraphics[width=1\linewidth]{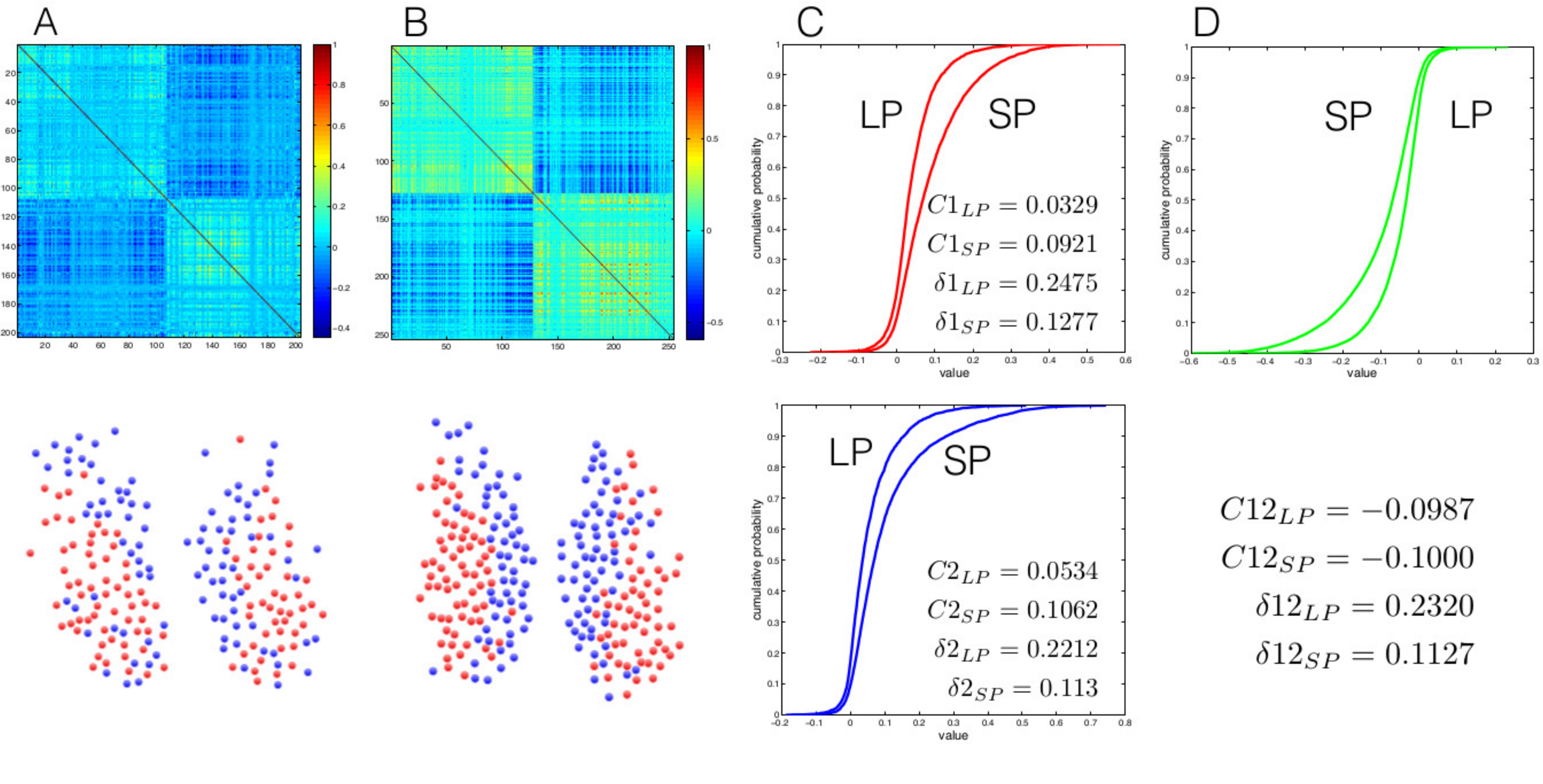} 
\caption{(A) The resolved functional signature and modules structure of a long photoperiod (LP, L16D8) sample. (B) The resolved functional signature and modules structure of a short photoperiod (SP, L8D16). (C) Cluster analysis: plotting the cumulative distribution of the dependencies within the two detected clusters, comparing the two different photoperiods. The upper graph shows cluster 1 and the lower graph cluster 2. $C$ represents the measured averaged correlation of a cluster, and $\delta \equiv \frac{N_{-}}{N_{+}}$ is defined as the contrast ratio of a cluster, measuring the ratio of negative dependencies versus positive dependencies. (D) Inter-cluster analysis: plotting the cumulative distribution of the external dependencies between the two detected clusters,comparing the two different photoperiods. $C$ represents the measured averaged anti-correlation between clusters, and $\delta \equiv \frac{N_{+}}{N_{-}}$ is defined as the contrast ratio of a cluster, measuring the ratio of positive dependencies versus negative dependencies.
}
\label{fig:extra}
\end{figure*}

Our method detects mostly two communities which coincide with the core and shell distinction within the SCN \cite{Morin}. The core of the SCN receives light input and adjusts quickly to changing light schemes, while the shell of the SCN lags behind \cite{Albus}. Mostly the core-shell distinction of the SCN is interpreted as a distinction between the ventrolateral and the dorsomedial part of the SCN, which is predominantly based on anatomical data \cite{Moore}. In this study the two clusters that were found were more dorsolaterally and ventromedially located, and while it is based on functional data this may differ from known anatomical distinctions. Furthermore, the SCN is much more heterogeneous when looked at cellular phenotype or gene expression \cite{Antle,Silver}. The anatomical loci do not necessarily delineate the phenotypical SCN regions very precisely, which implies that functionally, the core-shell distinction is less clearly defined and may differ from the described anatomical division (see also \cite{Morin}).

Regional analysis of the SCN using functional time series has been performed by other groups. Evans and co-workers used a similar approach to identify single-cell-like regions of interest, but did not use clustering algorithms and chose the regions by hand \cite{Evans2}. Silver and co-workers also used regions of interest, called superpixels, but these were not necessarily identified as single-cells. Based on these superpixels they used threshold methods to identify regional differences in the SCN \cite{Foley, Pauls}. Abel and co-workers applied a threshold method based on mutual information on single-cell-like regions of interest \cite{Abel}. These approaches encounter similar problems as described in this paper when using the threshold method: they only find one cluster (in the core, or ventral part) and many non-clustered cell-like ROIs (in the shell or dorsal part). Our results presented here are in line with the regional division of the SCN proposed in these studies, but we were able to identify both the core and shell clusters.

Furthermore, our approach is able to identify the two clusters in different experimental conditions, ranging from summer conditions (long days, short nights: LD 16h:8h) to winter conditions (short days, long nights: LD 8h:16h). On the contrary, Evans and co-workers identified changes occuring in the organization of the SCN, where the two regions similar to our clusters were found, only for very long day conditions (LD 20h:4h) \cite{Evans}.

As a next step we analyzed the values in the functional signatures, i.e. the filtered correlations, and compared them across different photoperiods that the animals have been subjected to. With this step we reveal the dynamics within the population of neurons in the clusters and between the clusters. 
As the cluster-partition is based on the functional signature, we can now investigate the values within and between the clusters, exploring the inner and outer level of correlation. This extra information links physiological properties of the SCN to the functional signature found in the data. We measure the average residual correlation within each cluster detected by our method and we plot the community distribution of the measured values [Fig~\ref{fig:extra}A,B]. We then identify the cumulative probabilty of the values in the clusters and we see that in short photoperiods the average values are much higher than in long photoperiods [Fig~\ref{fig:extra}C]. This means that the correlation within the clusters is significantly higher in short photoperiods than in long photoperiods. When we examine the values between the clusters, we see that the average value is lower in short versus long photoperiod, meaning that the clusters are less correlated in short photoperiods [Fig~\ref{fig:extra}D]. These results connect directly to previous results in physiological properties as described in \cite{Buijink}
and is supported in other papers \cite{VanderLeest,Rohling2}. Thus, we show that the hidden functional representation reveals the phase ordering of oscillating cell populations caused by physiological properties of the SCN.

\section*{Discussion}

Our method reveals hidden functional dependencies that are obfuscated by the presence of a global mode in the neuronal gene expression, which imparts an overall positive correlation.
This problem becomes particularly evident when searching for functional structure in neuronal systems where the global signal is very strong, making the identification of functional modules very challenging. 
Our method is able to deal with the joint effects of noise  and common global trends in the original data in a robust manner.  
In fact, we have shown that the effects of noise and those of the global signal are coupled, as their signatures in the spectrum of the correlation matrix depend on each other.

We found a distinctive left-right functional symmetry with core-shell features in the SCN. This structure reveals non-contiguous regions that display strongly synchronized activity, despite being at a relatively large distance from each other, similar to \cite{Nicosia}.
Remarkably, here we detect this functional symmetry on a micro-scale level where nodes are single cells.
In this respect, it is important to notice that while the traditional threshold method applied to the SCN resolves only a radial gradient of functional connectivity that closely mirrors the anatomical proximity of neurons without singling out any modularity or boundary, our method systematically reveals two sharp modules, a ventral core and a dorsal periphery. 
These modules feature distinct signatures of functional (as opposed to structural) connectivity, namely left-right symmetry, spatial non-contiguity, and almost perfect dynamical anti-correlation once the global SCN-wide signal is filtered out.
The left and right shell regions of the SCN, despite being spatially disconnected into two non-contiguous regions, are functionally joined into a single module.
These symmetrical structures in the SCN raise important questions with respect to the underlying mechanisms at work in the system, and can possibly be explored in the future.  

The ability to exploit all the information from the correlation matrix, i.e. both the negative and the positive dependencies (correlation and anti-correlation), in order to detect the functional modules is very powerful. 
The strength of our method is to detect communal phase differences in neuronal networks by analysing time series data without using any presumptions or threshold definitions. Phase differences and phase adjustments in neuronal networks are an key feature for physiological function and can be used to define the functional state of a network in health and disease. 
Our method allows the identification of synchronized clusters of cells. Synchronization within a neuronal network was suggested to play a major role in the occurrence of epilepsy \cite{Jiruska1,Jiruska2}, Parkinsons disease \cite{Lipski,Babiloni} and schizophrenia \cite{Uhlhaas1,Uhlhaas2}.
It is noteworthy that the clusters determined with our methods are not influenced by the functional change in E-I balance occurring in different photoperiods. This is advantegous since our analysis will also detect functional clusters within neuronal networks with altered E/I balance often found in neurological disease (e.g. epilepsy, RETT, FragileX, autism) and in the aging brain.

The results presented here show that our method offers great potential for detecting hidden functional synchronization and desynchronization in brain networks and are not limited to gene expression rhythms. Time series from other modalities, such as electrical action potential recordings, EEG recordings and fMRI recordings can also be interpreted through this new method. As such, the method may offer diagnostic or pre-diagnostic applications in medical health care.

\acknowledgements

This work was partially supported by the EU project MULTIPLEX (contract 317532) and the Netherlands Organization for Scientific Research (NWO/OCW).
AA and DG also acknowledge support from the Dutch Econophysics Foundation (Stichting Econophysics, Leiden, the Netherlands).
SM acknowledge support from the Netherlands Foundation of Technology [STW, ONTIME 12191]. \\

We thank Carlo Nicolini for stimulating discussions and for writing a code implementing our method. We thank Dr. Gabriella Lundkvist, Swedish Medical Nanoscience Center, Department of Neuroscience, Karolinska Institutet, for providing the PER2::LUC mice. We thank Dr. Henk-Tjebbe van der Leest and Trudy van Kempen for their contribution to the development of the bioluminescence imaging technique and analysis.

\section*{AUTHOR CONTRIBUTIONS}


Mathematical method: AA DG. Analyzed the data: AA OR JR DG. Wrote the manuscript: AA SM JR DG. Conceived and designed the experiments: SM JM. Performed the experiments: RB.

\section*{COMPETING FINANCIAL INTERESTS}
The authors declare no competing financial interests.

\end{document}